\documentclass[conference]{IEEEtran}
\IEEEoverridecommandlockouts
\usepackage{amsthm}
\usepackage{amsmath}
\usepackage{amssymb}
\usepackage{booktabs}
\usepackage{siunitx}
\usepackage{cite}
\usepackage{graphics, framed}
\usepackage{graphicx}
\usepackage{epsfig}
\usepackage{epstopdf}
\usepackage{subfigure}
\usepackage{url}
\usepackage{multimedia}
\usepackage{paralist}
\usepackage[printonlyused]{acronym}
\usepackage{units}
\usepackage{balance}
\usepackage{multirow}
\usepackage{tikz}
\usepackage{lipsum,adjustbox}
\usetikzlibrary{calc,positioning}
\usetikzlibrary{shapes}
\usetikzlibrary{arrows}
\usetikzlibrary{shapes.geometric, arrows}
\tikzstyle{startstop} = [rectangle, rounded corners, minimum width=1.75cm, minimum height=0.75cm,text centered, draw=black, fill=white]
\tikzstyle{io} = [trapezium, trapezium left angle=70, trapezium right angle=110, minimum width=1.75cm, minimum height=0.75cm, text centered, text width=1.75cm, draw=black, fill=white]
\tikzstyle{process} = [rectangle, minimum width=1.75cm, minimum height=0.75cm, text centered, text width=1.75cm, draw=black, fill=white]
\tikzstyle{decision} = [diamond, minimum width=1.75cm, minimum height=0.75cm, text centered, draw=black, fill=white]
\tikzstyle{arrow} = [thick,->,>=stealth]
\newcommand{\FGR}[1]{Fig.~\ref{#1}}

\newcommand{\SEC}[1]{Section~\ref{#1}}
\newcommand{\TAB}[1]{Table~\ref{#1}}

\acrodef{CFAR}[CFAR]{constant false alarm rate}
\acrodef{AWGN}[AWGN]{additive white Gaussian noise}
\acrodef{CR}[CR]{cognitive radio}
\acrodef{SNR}[SNR]{signal--to--noise ratio}
\acrodef{ADAM}[ADAM]{adaptive moment estimation}
\acrodef{ISM}[ISM]{industrial, scientific and medical}
\acrodef{CPU}[CPU]{central processing unit}
\acrodef{GPU}[GPU]{graphics processing unit}
\acrodef{LB}[LB]{likelihood based}
\acrodef{FB}[FB]{feature based}
\acrodef{AM}[AM]{amplitude modulation}
\acrodef{PSK}[PSK]{phase shift keying}
\acrodef{FSK}[FSK]{frequency shift keying}
\acrodef{QAM}[QAM]{quadrature amplitude modulation}
\acrodef{I/Q}[I/Q]{in--phase/quadrature}
\acrodef{AI}[AI]{artificial intelligence}
\acrodef{CNN}[CNN]{Convolutional neural network}
\acrodef{LSTM}[LSTM]{long short term memory}
\acrodef{DL}[DL]{deep learning}
\acrodef{CLDNN}[CLDNN]{convolutional long short term memory fully connected deep neural network}
\acrodef{5G}[5G]{5\textsuperscript{th}--Generation}
\acrodef{BW}[BW]{bandwidth}
\acrodef{CW}[CW]{continuous wave}
\acrodef{D2D}[D2D]{device--to--device}
\acrodef{M2M}[M2M]{machine--to--machine}
\acrodef{dB}[dB]{decibel}
\acrodef{dBi}[dBi]{decibel isotropic}
\acrodef{dBm}[dBm]{decibel over a milliwatt}
\acrodef{Gbps}[Gbps]{Gigabit per second}
\acrodef{GHz}[GHz]{gigahertz}
\acrodef{Hz}[Hz]{hertz}
\acrodef{IF}[IF]{intermediate frequency}
\acrodef{IFFT}[IFFT]{inverse fast Fourier Transform}
\acrodef{LO}[LO]{local oscillator}
\acrodef{LOS}[LOS]{line--of--sight}
\acrodef{MHz}[MHz]{megahertz}
\acrodef{MIMO}[MIMO]{multiple--input multiple--output}
\acrodef{mmWave}[mmWave]{millimeter wave}
\acrodef{NGWN}[NGWN]{next generation wireless network}
\acrodef{NLOS}[NLOS]{non line--of--sight}
\acrodef{QoS}[QoS]{quality of service}
\acrodef{RF}[RF]{radio frequency}
\acrodef{MLE}[MLE]{maximum likelihood estimation}
\acrodef{UHF}[UHF]{ultra high frequency}
\acrodef{AMC}[AMC]{Automatic modulation classification}
\acrodef{SCF}[SCF]{spectral correlation function}
\acrodef{SVM}[SVM]{support vector machine}
\acrodef{FFT}[FFT]{fast Fourier Transform}
\acrodef{RNN}[RNN]{recurrent neural network} 
\acrodef{ReLU}[ReLU]{rectified linear unit}
\acrodef{BS}[BS]{base station}
\acrodef{AP}[AP]{access point}
\acrodef{SON}[SON]{self organized network}
\acrodef{UAV}[UAV]{unmanned aerial vehicle}
\acrodef{LOS}[LOS]{line--of--sight}
\acrodef{NLOS}[NLOS]{nonline--of--sight}
\acrodef{OLOS}[OLOS]{obstructed line--of--sight}
\acrodef{IoT}[IoT]{Internet of Things}
\acrodef{I/Q}[I/Q]{in--phase/quadrature}
\ifCLASSINFOpdf
\else
\fi

\begin{document}
\title{Measurement Based Statistical Channel Characterization of Air--to--Ground Path Loss Model at 446MHz for Narrow--Band Signals in Low Altitude UAVs\thanks{This paper has been accepted for the presentation in the 2020 IEEE 91st Vehicular Technology Conference (VTC2020-Spring).}}
\IEEEoverridecommandlockouts 

\author{\IEEEauthorblockN{Burak Ede\IEEEauthorrefmark{1}\IEEEauthorrefmark{2}, Serhan Yarkan\IEEEauthorrefmark{1}\IEEEauthorrefmark{3}, Ali Rıza Ekti \IEEEauthorrefmark{1}\IEEEauthorrefmark{4}, Tun\c{c}er Bayka\c{s}\IEEEauthorrefmark{5}, Hakan Ali \c{C}{\i}rpan\IEEEauthorrefmark{2}, Ali G\"{o}r\c{c}in\IEEEauthorrefmark{1}\IEEEauthorrefmark{6}}
\IEEEauthorblockA{\IEEEauthorrefmark{1} Informatics and Information Security Research Center (B{\.{I}}LGEM), T{\"{U}}B{\.{I}}TAK, Kocaeli, Turkey}

\IEEEauthorblockA{\IEEEauthorrefmark{2} Department of Electronics and Communication Engineering, Istanbul Technical University, {\.{I}}stanbul, Turkey}

\IEEEauthorblockA{\IEEEauthorrefmark{3} Department of Electrical and Electronics Engineering, Istanbul Commerce University, {\.{I}}stanbul, Turkey}

\IEEEauthorblockA{\IEEEauthorrefmark{4} Department of Electrical and Electronics Engineering, Bal{{\i}}kesir University, Bal{{\i}}kesir, Turkey}

\IEEEauthorblockA{\IEEEauthorrefmark{5} Department of Electrical and  Electronics Engineering, Medipol University, {\.{I}}stanbul, Turkey}

\IEEEauthorblockA{\IEEEauthorrefmark{6} Faculty of Electronics and Communications Engineering, Y{{\i}}ld{{\i}}z Technical University, {\.{I}}stanbul, Turkey}\\ Emails: \texttt{burak.ede@tubitak.gov.tr,} \texttt{syarkan@ticaret.edu.tr,} \texttt{arekti@balikesir.edu.tr,}\\  \texttt{tbaykas@medipol.edu.tr,} \texttt{hakan.cirpan@itu.edu.tr,} \texttt{agorcin@yildiz.edu.tr}}

\maketitle

\begin{abstract}
Powered by the advances in microelectronics technologies, \acp{UAV} provide a vast variety of services ranging from surveillance to delivery in both military and civilian domains. It is clear that a successful operation in those services relies heavily on wireless communication technologies. Even though wireless communication techniques could be considered to reach a certain level of maturity, wireless communication links including \acp{UAV} should be regarded in a different way due to the peculiar characteristics of \acp{UAV} such as agility in $3$D spatial domain and versatility in modes of operation. Such mobility characteristics in a vast variety of environmental diversity render links including \acp{UAV} different from those in traditional, terrestrial mobility scenarios. Furthermore, \acp{UAV} are critical instruments for network operators in order to provide basic voice and short messaging services for narrow band communication in and around disaster areas. It is obvious that such widespread use of \acp{UAV} under different scenarios and environments requires a better understanding the behavior of the communication links that include \acp{UAV}. Therefore, in this study, details of a measurement campaign designed to collect data for large-scale propagation characterization of air--to--ground links operated by \acp{UAV} at 446MHz under narrowband assumption are given. Data collection, post-processing, and measurement results are provided.
\end{abstract}
\IEEEpeerreviewmaketitle
\acresetall

\section{Introduction}


Ubiquitous access has become an essential part of modern daily life. Services, applications, devices, and even products are equipped with wireless communication interfaces in such a way that diverse fields and domains are connected anywhere, anytime. In parallel with the escalating demand for high data rates, everywhere connectivity with mobility support becomes an indispensable design requirement for both contemporary and emerging wireless technologies. Although, it is expected that these wireless technologies should support crystal clear audio for voice, high--definition video downstreaming, very--low latency and $4$K online gaming, low-power consumption transmission, and so on, it is more important to provide service on mission critical scenarios. Obviously, such an aggressive set of requirements comes at the expense of several conflicting list of parameters. Furthermore, realization of these requirements necessitate utilization of crucial resources such as bandwidth and power. 

The conventional strategy to tackle the aforementioned utilization problem is reusing of resources as frequent as possible. Cell splitting, antenna sectoring, and small cell concepts are prominent examples to increase capacity via effective use of available bandwidth, reduce transmit power and impact of interference. There are products in the market regarding small cell (microcell, picocell, femtocell in the sense of metro femtocells, public access femtocells, enterprise femtocells and class $3$ level femtocells, \textit{etc.}) solutions which mainly rely on fixed deployment. However, with the emergence of \acp{UAV}, a paradigm shift has been experienced in wireless communication communities in many aspects. Considering its agility in $3$D spatial domain and versatility in modes of operation, \acp{UAV} bring about almost a completely novel perspective in contemporary wireless mobile radio communication systems. Ranging from emergency communications to routing and relaying, \acp{UAV} are considered to be promising solution candidates in various wireless communication scenarios. In addition, the emerging concept of \ac{IoT} mandates extending the wireless radio coverage to diverse (and relatively harsh propagation) environments. In the presence of such a vast variety of propagation environments, a successful wireless communication link depends on transceivers which rely on extended measurements and corresponding designs \cite{amorim_2017_path_loss}.

In the literature, there has been a significant attention in \acp{UAV} and measurement campaigns including several modes and scenarios. Considering the transmission modes akin to terrestrial ones, air--to--ground and air--to--air are the two prominent classes \cite{mozaffari_2015_drone}. Among these two, air--to--air class needs further investigation due to the aforementioned reasons relevant to \acp{UAV} modes of operations \cite{tu_2009_proposal, al_2017_modeling}. On the other hand, air-to-ground could be considered to be a transition class since it contains both terrestrial elements and \acp{UAV} simultaneously. Air-to-ground class consists of \ac{LOS}, \ac{NLOS}, and \ac{OLOS} \cite{yanmaz_2011_channel}. Of course, a comprehensive air--to--ground model requires the probabilistic transitional states for \ac{LOS}, \ac{NLOS}, and \ac{OLOS} cases to be defined as well \cite{feng_2006_path, feng_2006_wlcp2, bor_2017_environment}. It is obvious that further analysis is required to have an extended model which takes into account shadowing as an additional parameter \cite{weiner_1986_use, holis_2008_elevation, zeng_2017_second}. A very detailed collection of studies present in the literature could be found in \cite{vinogradov_2019_tutorial}. Beside statistical models which depend heavily on theoretical derivations \cite{simunek_2013_uav}, there are exact \cite{al_2014_optimal} and numerical approaches in determining the propagation characteristics as well. Especially ray tracing method that runs in downtown scenarios  with the extension of building heights is employed very frequently \cite{al_2014_modeling,al_2017_modeling}.

In the mission critical scenarios, where the utilization of \ac{UAV} \acp{BS}/\acp{AP} such as disaster/public safety regions, rural areas and downtown areas where the total failure of communication infrastructure would lead to catastrophic events in terms of wireless communication, finding a rapid and cost--effective recovery solution that utilizes narrow band voice channels will be an important task. There have been some commercial services offered by cellular operators which claim to employ drones as hovering base stations in order to provide coverage in emergency scenarios and/or when disaster strikes. However, one of such operators which claimed to have \ac{UAV} base stations could not provide service for a while after an earthquake of magnitude of 5.8 hit Istanbul, Turkey, on September 27th, 2020 \cite{dronecell}. This instance revealed that establishing communication via  \acp{UAV} in a disaster scenario is a challenging task which has multiple dimensions including careful propagation channel analysis, detailed network planning, and spectrum management for first responders \cite{ecc_roadmap,del2019radio, sanchoyerto2019analysis}. 

Despite all of the measurement campaigns, results, and theoretical analyses present in the literature which are mostly focused on 800MHz--850MHz, 960MHz--977MHz and 5030MHz--5091MHz for the \ac{UAV} communication, a generic and comprehensive propagation model for emergency case communications focusing specifically on \acp{UAV} operating at UHF bands is still required. Therefore, in this study, a single-frequency path loss measurement campaign at 446MHz is proposed. Air-to-ground link is established via a narrow band signal generator (NBSG) mounted on \ac{UAV}. Measurements are collected within a suburb district, which is located close by to an inner sea. Measurement site is of hilly terrain structure with foliage. Considering the fact that such topographical layouts are difficult to reach and collect measurement data, this study paves the way for incorporating various terrain profiles into the channel models and obtaining a more comprehensive air–-to-–ground propagation link. The organization of the paper is as follows: General characteristics of air--to--ground path loss link is presented in \SEC{sec_general_model}. In \SEC{sec_experimental_setup}, the details of the experimental setup, equipment, and data collection stages are outlined. Results and relevant discussions are presented in \SEC{sec_experimental_results}. Finally, conclusions are drawn in the last section.

\section{General Large Scale Characteristics of Air--to--Ground Link for UAV}
\label{sec_general_model}

In a general air--to--ground wireless communication link scenario, it is frequently reported in the literature that the propagation mechanism at large scale is governed by the following statistical path loss equation:

\begin{equation}
\Lambda(d)=\underbrace{20\times \log\frac{4\pi d_{0}}{\lambda}}_{PL_{0}} + 10 \eta \log \left(d / d_{0}\right),
\end{equation}

\noindent where $d$ is the transmitter-receiver separation, $\eta$ is the path loss coefficient which depends on the environment, and $d_{0}$ is the reference distance for path loss measurements along with $PL_{0}$ being the free--space offset or intercept. A further improvement is possible by incorporating the \ac{LOS} into the model as

\begin{equation}
\Lambda(d)=\underbrace{20\times \log\frac{4\pi d_{0}}{\lambda}}_{PL_{0}} + 10 \eta \log \left(d / d_{0}\right) + \mu_{LOS},
\end{equation}

\noindent where $\mu_{LOS}$ includes all of the large--scale characteristic losses for \ac{LOS}.



\section{Measurement Campaign}\label{sec_experimental_setup}
\subsection{Measurement Campaign and Measurement Setup}
The measurements are taken in the test field which is shown in \FGR{fig:dji_wind4} of The Scientific and Technological Research Council of Turkey (TUBITAK) in Gebze. A narrow band signal generator is used to transmit a single tone in 446MHz center frequency on the UAV side. DJI Wind 4 is flied up to obtain measurement and it is used as a air station since its payload capability and durability in extreme conditions. It is shown in \FGR{fig:dji_wind4}. 

\begin{figure*}[t!]
\centering
\begin{minipage}[b]{.45\textwidth}
\begin{subfigure}[Block diagram for measurement setup.]{\includegraphics[width=3.4in,height=2in]{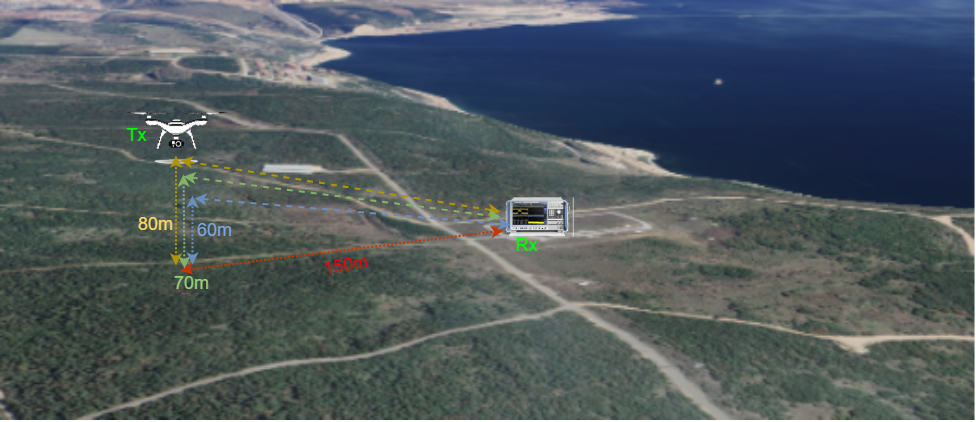}
\label{fig:dji_wind4}
}%
\end{subfigure}
\end{minipage}\qquad
\begin{minipage}[b]{.45\textwidth}
\begin{subfigure}[Rohde Schwarz FSW26 as receiver.]{\includegraphics[width=3.2 in,height=2in]{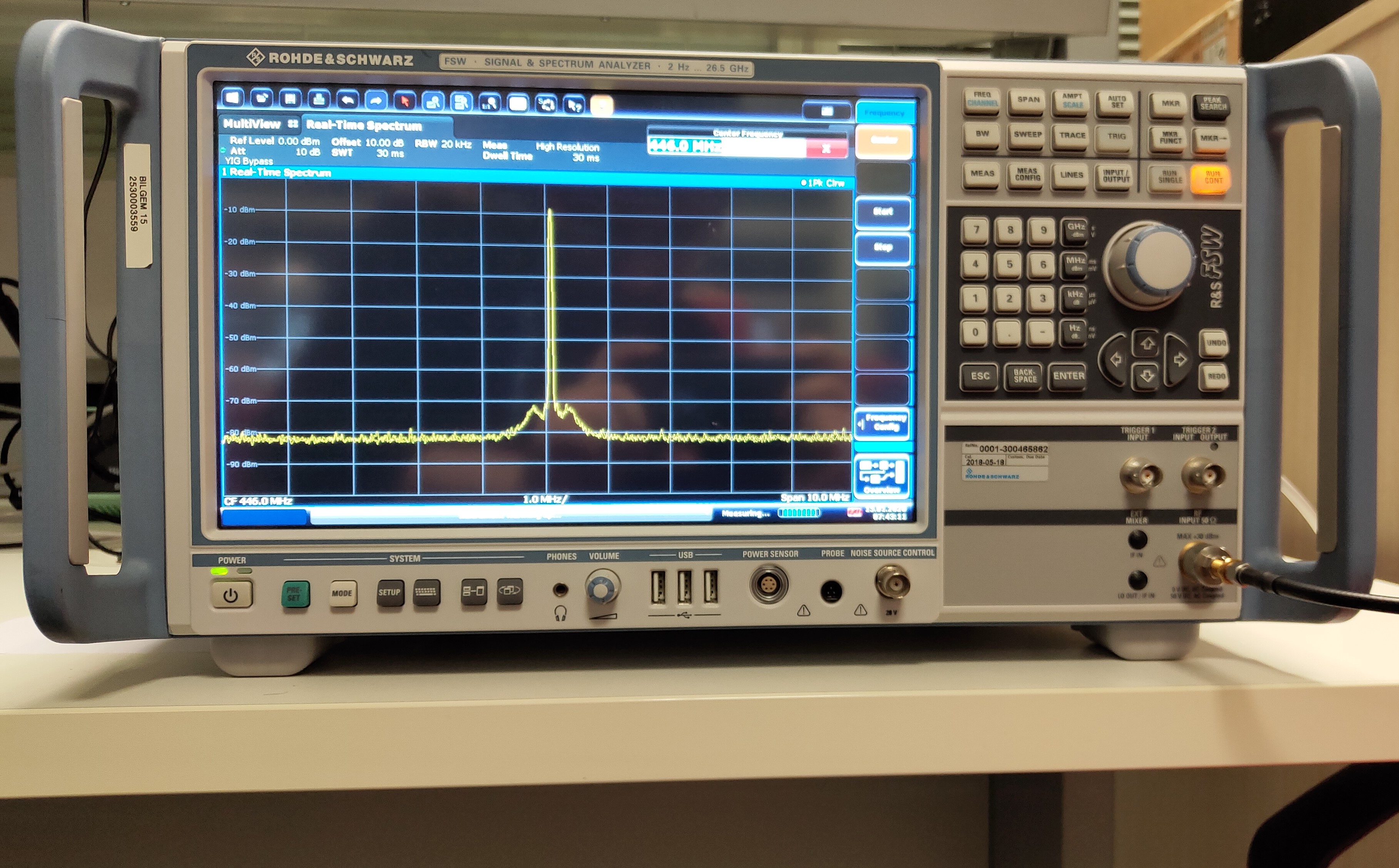}
\label{fig:rsw26}
}%
\end{subfigure}%
\end{minipage}
\caption{Measurement campaign and equipment.}
\label{fig:meas_camp_equipments}
\end{figure*}


On the other hand, R\&S FSW Signal and Spectrum Analyzer in \FGR{fig:rsw26}, is used as a ground station to receive the transmitting signals. It works in frequency range from 2Hz to 26.5GHz and offers up to 500MHz analysis bandwidth for measuring wideband--modulated or frequency--agile signals. Signals are sampled by 1MHz and they are recorded with 1sn duration. 
Furthermore, a Bowtie antenna which frequency range from 400MHz to 1000MHz is used for ground station to receive narrow band signals. The antenna has omnidirectional radiation pattern and it has 0.9dBi antenna gain at 400MHz. After the calculation of received powers of each point, then path loss model are extracted as it will be explained in \SEC{sec_experimental_results}. 
\subsection{Measurement Methodology}

All the field measurements are conducted in 446 MHz frequency band by using narrow band signal generator attached to a commercial \ac{UAV} and all the signals are captured by the ground station with 1MHz sample rate during one second. The measurements are collected from 60m to 80m heights with step size 10m vertically. Local regulations allow \ac{UAV} to operate up to 120m in height, however, due to the topographical conditions and the high wind speed after 80m in height, measurements are performend for 60m to 80m in height. Moreover, the only allocated location for \ac{UAV} flight is, where the measurement is taken, is shown \FGR{fig:dji_wind4} in TUBITAK, Gebze. Also, for each vertical measurement point, \ac{UAV} move between 50m to 200m horizontally. Overall grid consists of total 45 determined locations points with 60m to 80m vertically and 50m to 200m horizontally. Collected signals are fed into a computer that runs MATLAB 2015b. Raw time data which is given in \ac{I/Q} format are passed through a 255th order FIR filter to maximize power of narrow band signal and to get rid of spurious signals from the captured signal. In addition to this, the center frequency offset is calculated for each point and shifted by that value. Also, another FIR filter with 127th order is used for removing noise arising from harmonics. Furthermore, offset in the time domain, is calculated and eliminated from the processed signal. Then, the signal is down sampled to avoid complexity of the calculation and the average power with dB scale is calculated for each data set.

Measurement campaign workflow for the aforementioned methods is shown \FGR{fig:meas_work_flow}.
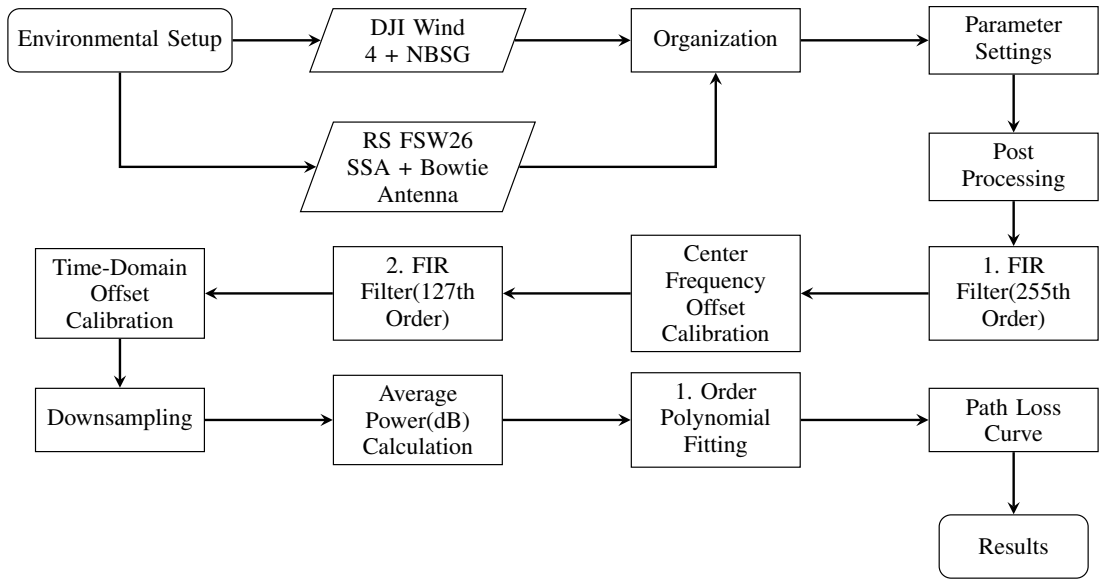
\begin {figure*}
\centering
\begin{adjustbox}{width=0.8\textwidth, height=3in}
\begin{tikzpicture}[node distance=1.5cm,scale=0.5]
    \tikzstyle{every node}=[font=\footnotesize]
    \node (start) [startstop] {Environmental Setup};
    \node (in1) [io, right of=start,xshift=2cm] {DJI Wind 4 + NBSG};
    \node (in2) [io, below of=in1] {RS FSW26 SSA + Bowtie Antenna};
    \node (pro1) [process, right of=in1,xshift=2cm] {Organization};
    \node (pro2) [process, right of=pro1,xshift=2cm] {Parameter Settings};
    \node (pro3) [process, below of=pro2] {Post Processing};
    \node (pro4) [process, below of=pro3] {1. FIR Filter(255th Order)};
    \node (pro5) [process, left of=pro4, xshift=-2cm] {Center Frequency Offset Calibration};
    \node (pro6) [process, left of=pro5, xshift=-2cm] {2. FIR Filter(127th Order)};
    \node (pro7) [process, left of=pro6, xshift=-2cm] {Time-Domain Offset Calibration};
    \node (pro8) [process, below of=pro7] {Downsampling};
    \node (pro9) [process, right of=pro8, xshift=2cm] {Average Power(dB) Calculation};
    \node (pro10) [process, right of=pro9, xshift=2cm] {1. Order Polynomial Fitting};
    \node (pro11) [process, right of=pro10, xshift=2cm] {Path Loss Curve};
    \node (stop) [startstop, below of=pro11] {Results};
    \draw [arrow] (start) -- (in1);
    \draw [arrow] (start) |-(in2);
    \draw [arrow] (in1) -- (pro1);
    \draw [arrow] (in2) -| (pro1);
    \draw [arrow] (pro1) -- (pro2);
    \draw [arrow] (pro2) -- (pro3);
    \draw [arrow] (pro3) -- (pro4);
    \draw [arrow] (pro4) -- (pro5);
    \draw [arrow] (pro5) -- (pro6);
    \draw [arrow] (pro6) -- (pro7);
    \draw [arrow] (pro7) -- (pro8);
    \draw [arrow] (pro8) -- (pro9);
    \draw [arrow] (pro9) -- (pro10);
    \draw [arrow] (pro10) -- (pro11);
    \draw [arrow] (pro11) -- (stop);
\end{tikzpicture}
\end{adjustbox}
\caption{Illustration of the measurement set--up and post--processing workflow.}
\label{fig:meas_work_flow}
\end{figure*}

\section{Measurement Results}\label{sec_experimental_results}
In this study, measurement campaign focuses in the following aspect on how the narrow band signal received power is related to the transmitter–-receiver separation by considering both horizontal and vertical distance. To observe this simultaneously, \FGR{fig:power_alt_vs_dist_all} is plotted. As seen in \FGR{fig:power_alt_vs_dist_all}, the received power decreases with the transmitter-–receiver separation, as expected. Based on the measurement data when the least-–squares estimation is applied, the following path loss models are obtained: 

\begin{equation}
    \text{PL}(\text{dB})=\left\{\begin{matrix}
-2.176\times d - 35.99,~a=60m\\ 
-1.244\times d - 55.23,~a=70m\\ -1.473\times d - 49.4,~a=80m
\end{matrix}\right.
\end{equation}
\noindent where $d$ is the transmitter--receiver separation and $a$ stands for the altitude of UAV.

For the sake of brevity, the path loss coefficients for specific frequency and altitudes are listed in \TAB{tbl_path_loss_exponents_for_specific}. Overall mean path loss exponent is found to be $\eta=2.176, 1.244, 1.473$ for the altitudes of \unit{60}{m}, \unit{70}{m}, and \unit{80}{m}, respectively. Another important observation, which is also in conformity with the results presented in the literature in different frequency \cite{cai2018empirical}, is that as the \ac{UAV} moves up in the horizontal direction, the path loss exponent, $\eta$, values decrease in general. Even though, this result looks surprising it is still in parallel with the independent measurement in some other studies in literature \cite{cai2018empirical}. This can be explained as the Tx--Rx communication link will be less sensitive to horizontal distance at high altitude and the channel condition is better in higher altitudes. Furthermore, the \ac{UAV} at high altitude is most likely be affected less in terms of received power change due to the fact that it flies in the direction of the non--dominant radiation pattern of the BS antenna. Thus, not only environment but also height dependent channel propagation parameters are needed for describing the propagation channel for \ac{UAV} to ground station communication links. 

\begin{table}[t!]
\centering
\caption{Values for $D_{\max}$ and Path Loss Exponent, $\eta$, for \unit{446}{MHz} at different altitudes.}
\label{tbl_path_loss_exponents_for_specific}
\begin{tabular}{|c|c|c|c|}
\hline
Altitude (m) & 60 & 70 & 80 \\ \hline
PL. Exp. ($\eta$) & 2.176   & 1.244   & 1.473   \\ \hline
\end{tabular}
\end{table}

\begin{figure*}[t!]
\centering
\begin{minipage}[b]{.3\textwidth}
\begin{subfigure}[60m altitude.]{\includegraphics[width=2.2in,height=2in]{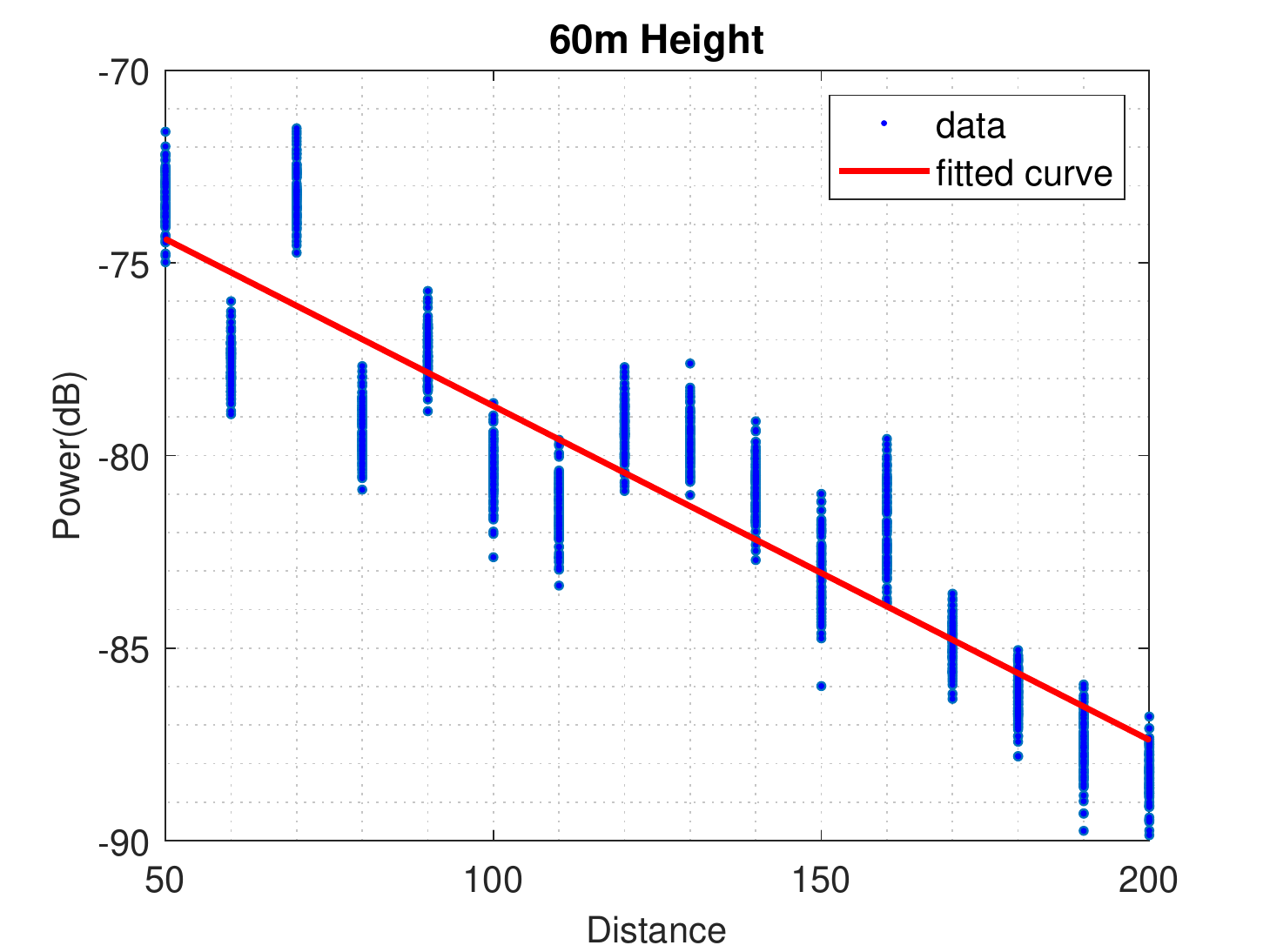}
\label{fig:60m_alt}
}%
\end{subfigure}
\end{minipage}\qquad
\begin{minipage}[b]{.3\textwidth}
\begin{subfigure}[70m altitude.]{\includegraphics[width=2.2in,height=2in]{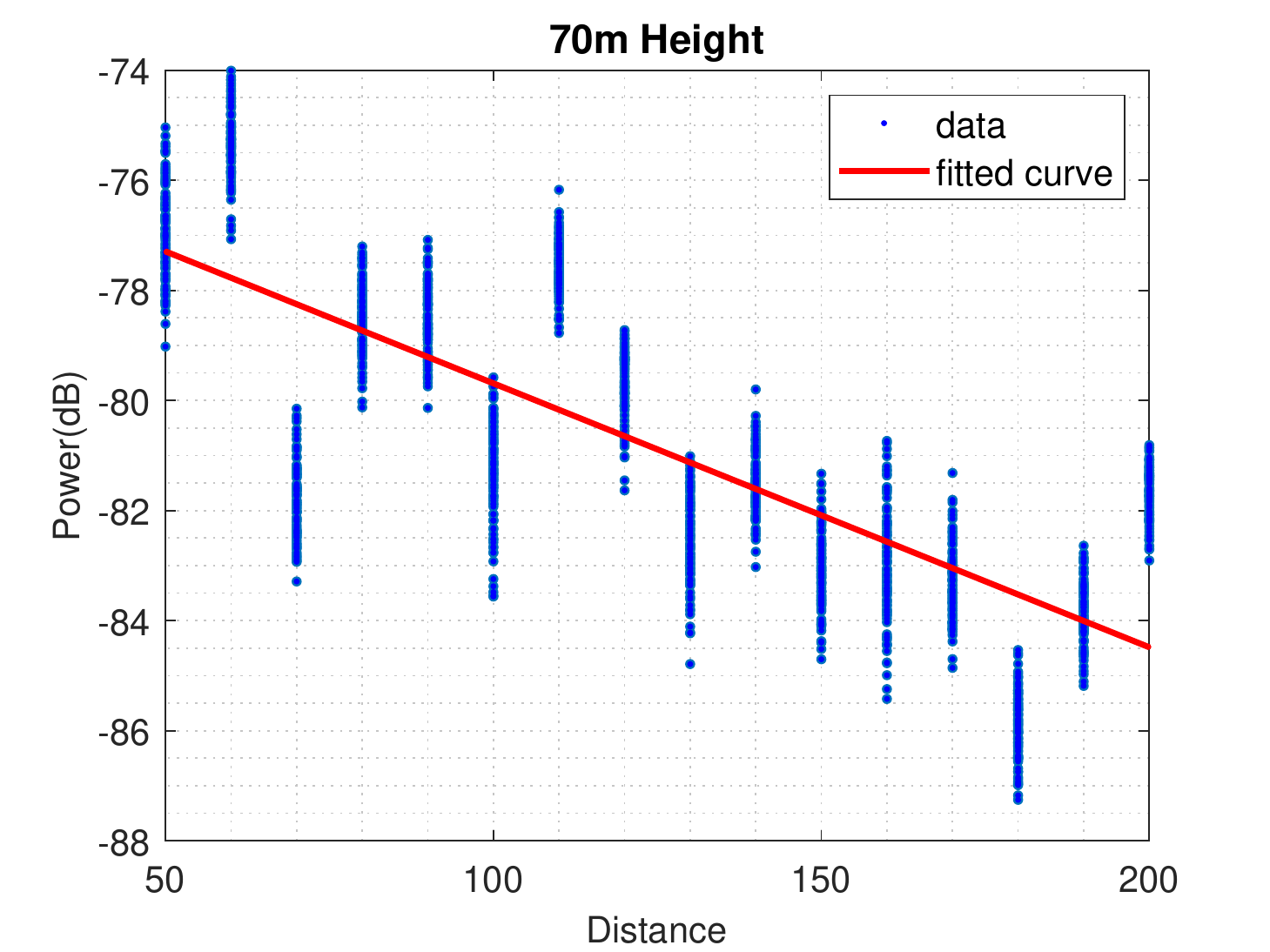}
\label{fig:70m_alt}
}%
\end{subfigure}%
\end{minipage}\qquad
\begin{minipage}[b]{.3\textwidth}
\begin{subfigure}[80m altitude.]{\includegraphics[width=2.2in,height=2in]{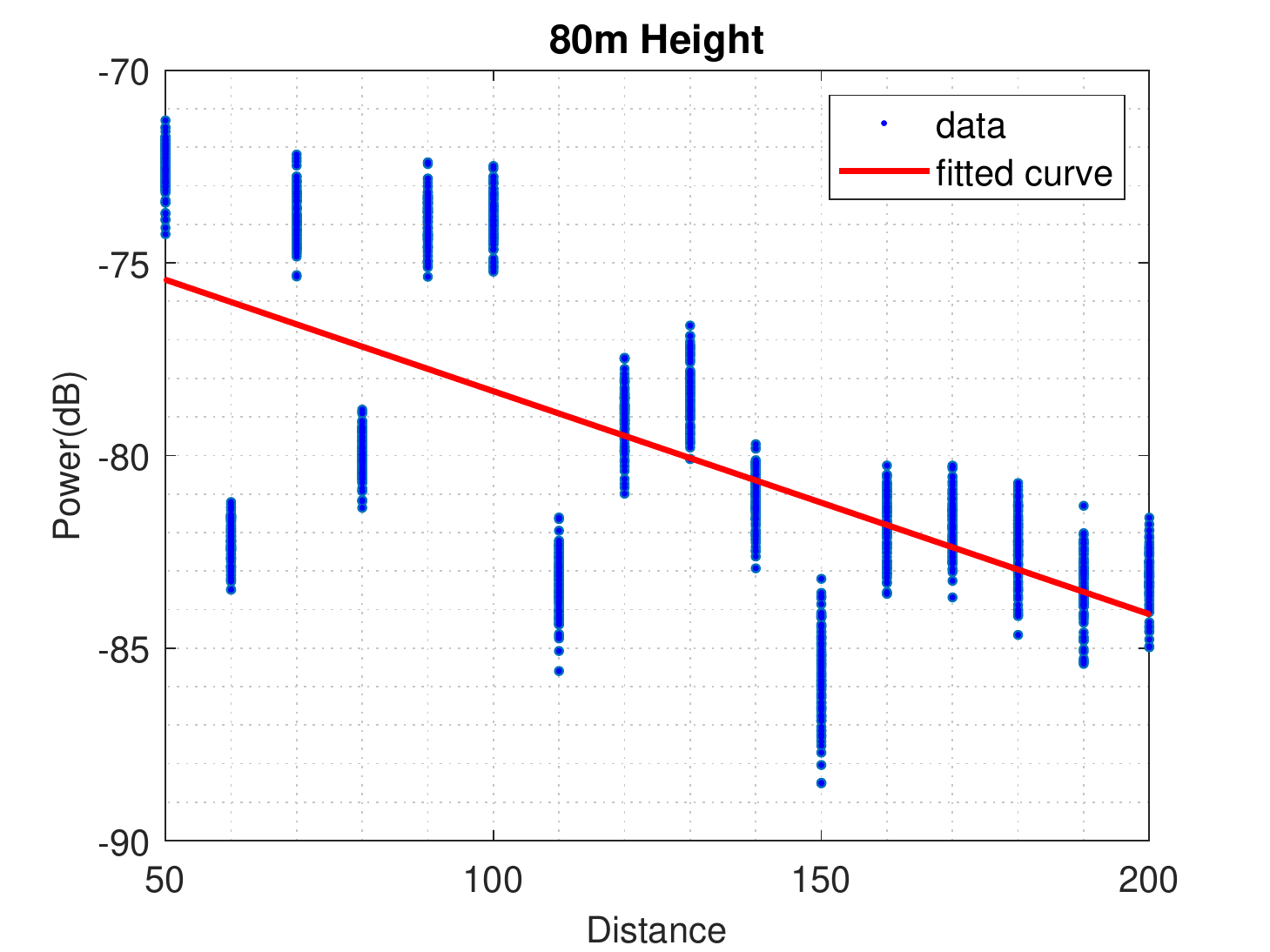}
\label{fig:80m_alt}
}%
\end{subfigure}%
\end{minipage}\qquad
\begin{minipage}[b]{.3\textwidth}
\begin{subfigure}[Comparison of PL curves for different heights.]{\includegraphics[width=2.2in,height=2in]{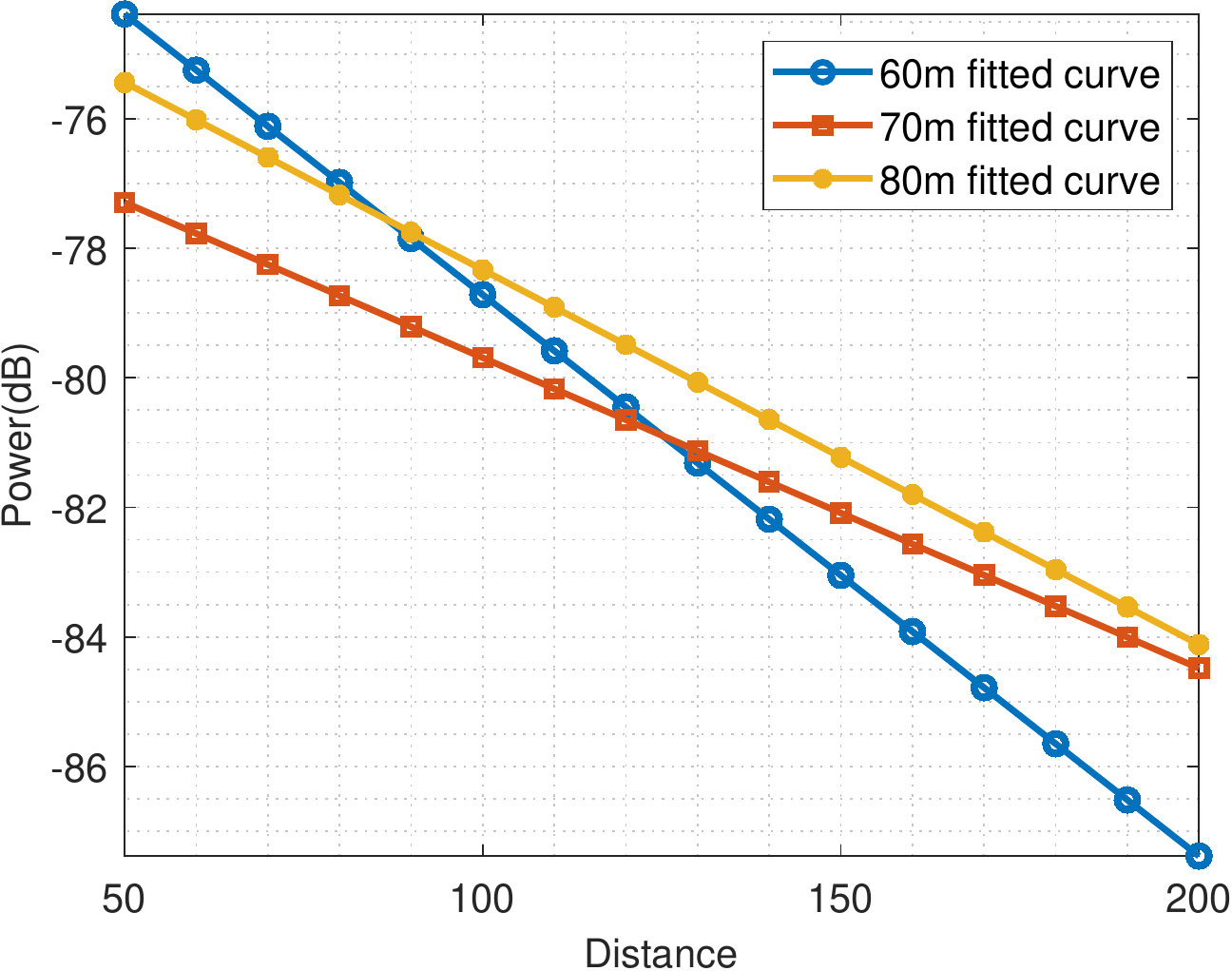}
\label{fig:compare}
}%
\end{subfigure}%
\end{minipage}\qquad
\caption{Measurement results for received power levels at different distance.}
\label{fig:power_alt_vs_dist_all}
\end{figure*}

\section{Concluding Remarks and Future Directions}\label{sec:conclusion}

In this work, the results of a measurement campaign for air to ground channels of \ac{UAV} at 446MHz are presented due to the fact that refarming of UHF band will enable new opportunities for emergency case situations. Based on the measurement data, it is shown that altitude along with the environment is an important factor that needs to be considered. Due to the increase in the elevation, the channel condition gets better and path loss exponent tends to decrease in general. Thus, in the future studies, we will extend our results to multiple frequency ranges by also incorporating environment affect in different locations and topographies.

\section*{Acknowledgement}
This work of Tuncer Baykas was supported by the Scientific and Techno	logical Research Council of Turkey (TUBITAK) under Grant 215E324.\\

Also, this publication was made possible by NPRP12S-0225-190152 from the Qatar National Research Fund (a member of The Qatar Foundation). The statements made herein are solely the responsibility of the author[s].
\balance
\bibliographystyle{IEEEtran}
\bibliography{VTC_Final_V2}
\end{document}